\documentclass[10pt, a4paper, twocolumn]{article} 

\usepackage[english]{babel} 

\usepackage{microtype} 

\usepackage{amsmath,amsfonts,amsthm} 

\usepackage[svgnames]{xcolor} 

\usepackage{booktabs} 

\usepackage{lastpage} 

\usepackage{enumitem} 
\setlist{noitemsep} 

\usepackage{sectsty} 
\allsectionsfont{\usefont{OT1}{phv}{b}{n}} 

\usepackage{geometry} 

\usepackage[T1]{fontenc} 
\usepackage[utf8]{inputenc} 

\usepackage{XCharter} 

\usepackage{titlesec} 
\renewcommand\thesection{\Roman{section}} 
\renewcommand\thesubsection{\roman{subsection}} 
\titleformat{\section}[block]{\large\scshape\centering}{\thesection.}{1em}{} 
\titleformat{\subsection}[block]{\large}{\thesubsection.}{1em}{} 

\usepackage{fancyhdr} 
\fancypagestyle{firstpage}{ 
	\fancyhf{}
}

\newcommand{\authorstyle}[1]{{\large\usefont{OT1}{phv}{b}{n}\color{DarkRed}#1}} 

\newcommand{\institution}[1]{{\footnotesize\usefont{OT1}{phv}{m}{sl}\color{Black}#1}} 

\usepackage{titling} 

\newcommand{\HorRule}{\color{DarkGoldenrod}\rule{\linewidth}{1pt}} 

\pretitle{
	\vspace{-30pt} 
	\HorRule\vspace{10pt} 
	\fontsize{22}{28}\usefont{OT1}{phv}{b}{n}\selectfont 
	\color{Black} 
}

\posttitle{\par\vskip 15pt} 

\preauthor{} 

\postauthor{ 

	\today
	\vspace{5pt} 
	\par\HorRule 
	\vspace{1pt} 
}

\usepackage{lettrine} 
\usepackage{fix-cm}	

\usepackage{xstring} 

\usepackage{graphicx}
\usepackage{float}
\usepackage{siunitx}
\usepackage{subcaption}
\usepackage{hyperref}
\hypersetup{
	colorlinks   = true, 
	urlcolor     = blue, 
	linkcolor    = blue, 
	citecolor   = red 
}
\usepackage[backend=biber, style=chem-angew]{biblatex}    
   
\title{A frequency quintupled laser at 308\,nm for spectroscopy of intercombination lines in zinc}
\author{
	\authorstyle{Maya Büki\textsuperscript{1, *}, David Röser\textsuperscript{1} and Simon Stellmer\textsuperscript{1}} 
	\newline\newline 
	\textsuperscript{1}\institution{University of Bonn, Nussallee 12, 53115 Bonn, Germany}\\ 
	\textsuperscript{*}\institution{Corresponding author: maya.bueki@physik.uni-bonn.de}\\ 
	\vspace{5pt}
}

 \date{\vspace{-5ex}}

\begin{document}

\maketitle 

\thispagestyle{firstpage} 


\noindent
\textbf{Many experiments in atomic physics and quantum optics, among them optical atomic clocks, require laser sources in the ultra-violet wavelength range with very low intensity noise and phase noise. The development of such lasers is a challenge, especially when a robust and transportable system is required. Here, we report on the development of a novel continuous wave (cw) frequency-quintupled laser at 308\,nm with an output power of 0.5\,mW, based on a fiber laser operating in the telecom band. Three consecutive frequency conversion stages in nonlinear crystals are employed. The performance of the laser system is demonstrated by linear absorption spectroscopy of a narrow intercombination line in zinc.}


\section{Introduction}
High-power, continuous-wave (cw), and narrow-linewidth laser sources are available only at visible and infrared (IR) wavelengths. To generate such radiation at UV wavelengths, nonlinear frequency conversion has been tremendously successful. 
Fiber lasers and fiber amplifiers combine high power output with very low intensity and phase noise, which are highly desired properties for spectroscopy and atomic physics experiments, including laser cooling and optical clocks. The output wavelength of fiber laser is limited to the IR spectrum: roughly 1030 to 1120\,nm for Yb-doped fibers, 1535 to 1580\,nm for Er-doped models, and 1900 to 2100\,nm for Tm-doped models. Most relevant atomic transitions, however, appear at UV and visible wavelengths.

Here, we report on the development of a frequency-quintupled Er-doped fiber laser to access the wavelength range of about 307 to 316\,nm. This wavelength range covers a range of transitions that are relevant for laser cooling and optical clocks, including the cooling transitions of Be$^+$ ions (313\,nm) and neutral Zn atoms (308\,nm), the clock transition in Zn (310\,nm), and the direct Rydberg excitation in Cs (318\,nm). It also includes the well-established 308-nm emission wavelength of XeCl excimer lasers.

Our setup does not involve optical cavities, which would increase the conversion efficiencies, but also add considerable complexity to the system. Instead, we start with a high-power fundamental laser source and employ single-pass frequency conversion. As a result, the setup described here requires only a small footprint and is highly robust. These assets, combined with the performance of well-established fiber laser technology, are highly beneficial for the applications mentioned above. This applies especially to transportable laser systems. Our scheme of generating cw output at the odd harmonics could readily be adapted to fiber laser models with other fundamental wavelengths \cite{Hansen2015}.

\section{A frequency quintupled laser at 308 nm}
\begin{figure}[ht!]
	\centering
	\includegraphics[width=\linewidth]{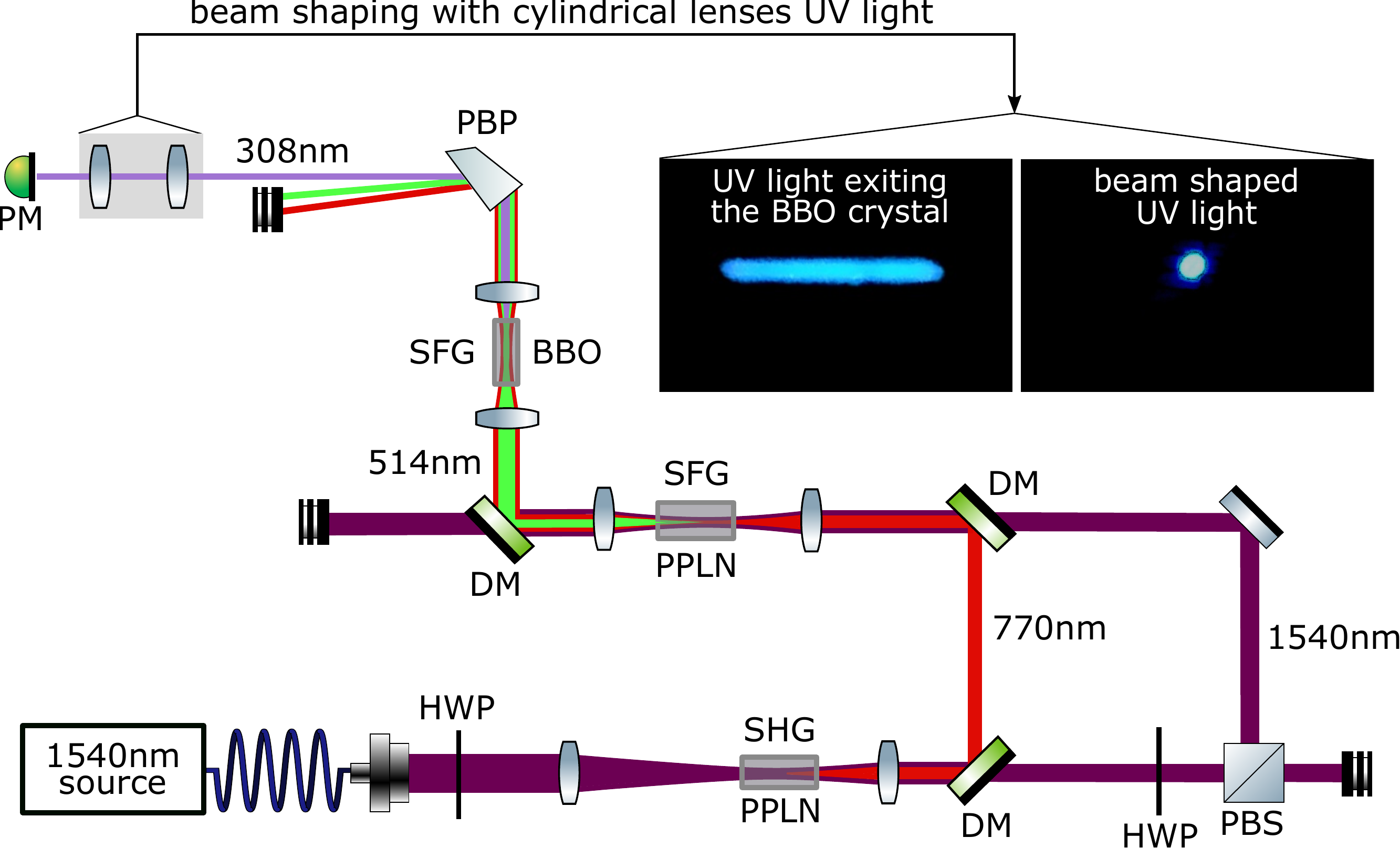}
	\caption{Schematic of the frequency quintupling laser setup with three conversion stages. First stage: Second harmonic generation (SHG) producing red light at \SI{770}{nm} using a periodically poled lithium niobate crystal (PPLN). Second stage: Sum frequency generation (SFG) producing green light at \SI{514}{nm} using PPLN. Third stage: SFG producing UV light at \SI{308}{nm} using a beta barium borate crystal (BBO). The power of the infrared light is reduced using a half wave plate (HWP) and a polarising beam splitter (PBS) in order to minimise green induced infra red absorption (GRIIRA). DM = dichroic mirror, PBP = Pellin-Broca-prism, PM = power meter. The inset shows the beam profile before and after beam shaping with cylindrical lenses.}
	\label{fig:setup}
\end{figure}

The setup is based on a fundamental laser source operating in the telecom C band (1530 to 1565\,nm). Two different types of lasers are employed: a distributed feedback diode laser with a linewidth of less than 1\,MHz and a tuning range of 2.5\,nm (Toptica Photonics AG), and a fiber laser with much smaller linewidth ($< \SI{100}{Hz}$) and a smaller tuning range of 1\,nm (Koheras ADJUSTIK from NKT Photonics).
Both lasers emit a few 10\,mW of power, a fraction of which is required to seed an Er-doped fiber amplifier (Koheras BOOSTIK HP from NKT Photonics) that increases the power level to up to 17\,W over a wide wavelength range without significant linewidth broadening. The two seed lasers operate at slightly different center wavelengths (1540 and 1550\,nm) and are easily interchangeable.
Compared to pulsed lasers, the peak power of cw lasers is low, and direct generation of high harmonics is very inefficient. Instead, we employ a cascade of three low-order processes. The last conversion process, which generates the UV light at five times the fundamental frequency $\nu$, is the one of least efficiency. Here, two interactions are available: $2\nu + 3\nu = 5\nu$ and $1\nu + 4\nu = 5\nu$.
In the following, we describe the generation of the second, third, and fifth harmonic in detail.

In order to use periodically-poled 5\%-MgO-doped lithium niobate (MgO:PPLN) for the first two conversion stages, we opt for the variant of $2\nu + 3\nu = 5\nu$ as MgO:PPLN is non-transparent for $4\nu$. The reasoning in choosing MgO:PPLN lies in its effective nonlinear conversion coefficient of typically 14 to 16\,\si{pm/V} for a Type 0 phase matching configuration, which is the highest coefficient in a single pass configuration compared to other nonlinear materials.
For efficient frequency conversion, we optimise the polarisation, beam alignment, crystal temperature, and beam waist. One advantage of the proposed setup as shown in Fig.~\ref{fig:setup} is that the polarisation of the light needs to be adjusted only once before entering the first crystal. Throughout the first two conversion stages, the fundamental and frequency-converted lights exhibit the same polarisation (Type 0 phase matching). In the first two stages, the temperatures of the crystals (MSHG1550-0.5-40 and MOPO515-0.5-40, provided by Covesion Ltd.) are controlled to within $\pm\SI{0.1}{\degreeCelsius}$ in order to achieve stable phase matching. To this end, the crystals are placed in temperature stabilized ovens, which in turn are mounted on 5-axis-stages for precision alignment.

\begin{figure*}[!hbt]
	\centering
	\includegraphics[width=1.0\linewidth]{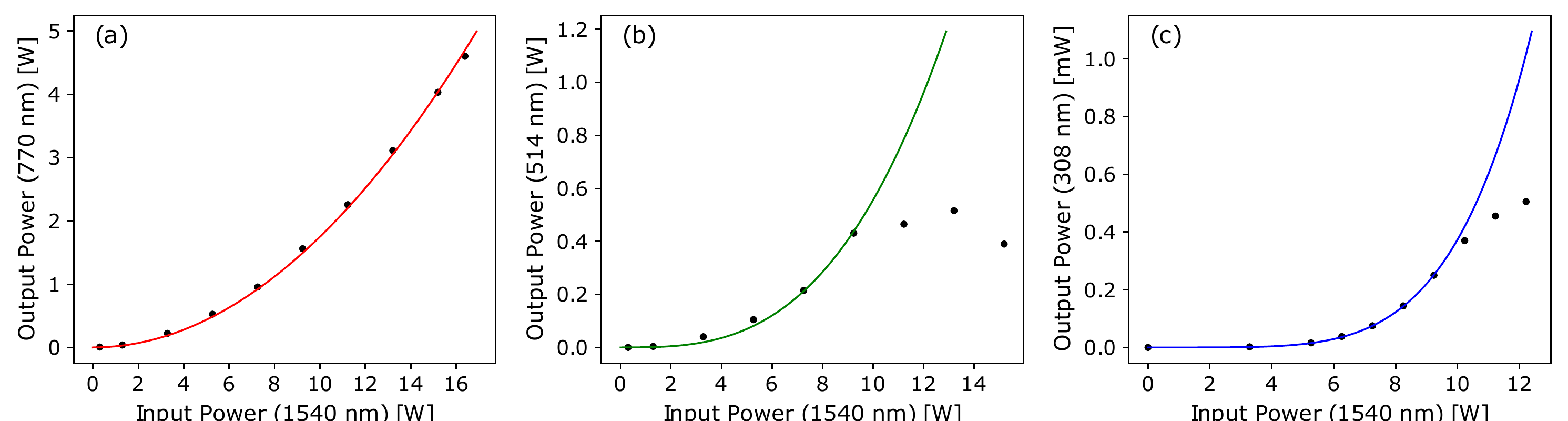}
	\caption{Measured power of the frequency converted light in the first, second and third conversion stage as a function of total infrared power (black dots). The solid curves represent polynomial fits to the data, with powers of (a) $P_{\rm IR}^2$, (b) $P_{\rm IR}^3$, and (c) $P_{\rm IR}^5$.}
	\label{fig:power}
\end{figure*}

The MgO:PPLN crystal for the first SHG stage has a length of 40\,mm, a poling period of \SI{19.1}{\micro\meter}, and is kept at a temperature of around \SI{115}{\degreeCelsius}. The 1540-nm light is focused to a waist of \SI{62}{\micro\meter} inside the crystal, following the theory of Boyd and Kleinman \cite{Boyd}. A conversion efficiency of \SI{0.44}{\%/W cm} returns almost 5\,W at 770\,nm; see Fig.~\ref{fig:power}(a).

\begin{figure}[b]
	\centering
	\includegraphics[width=0.8\linewidth]{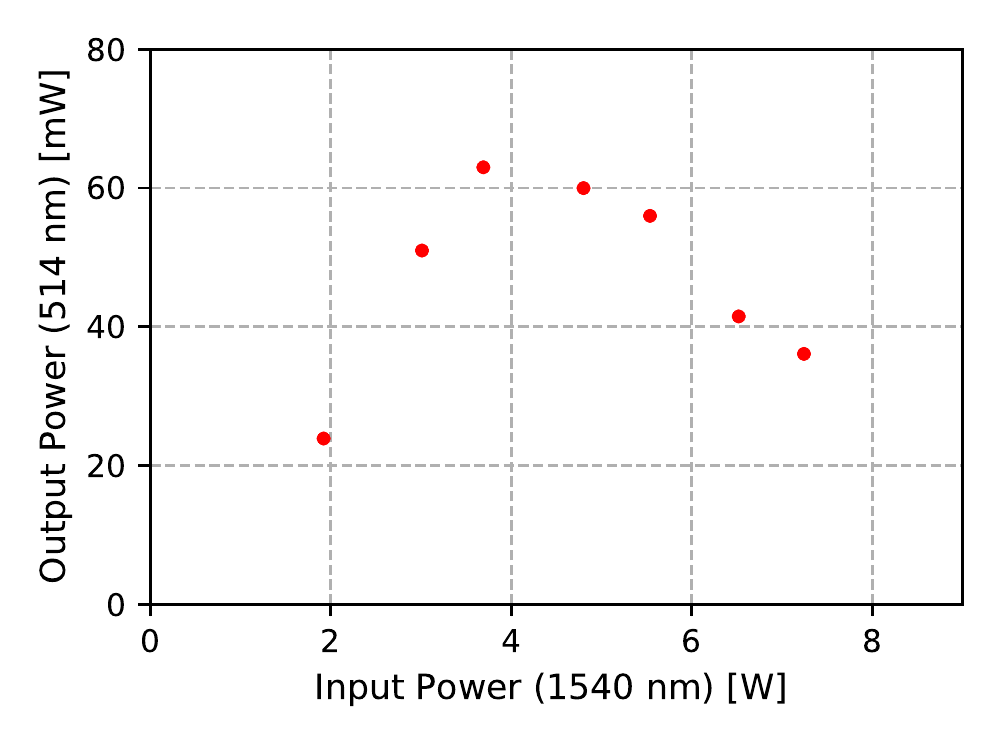}
	\caption{Measured output power at \SI{514}{\nano \meter} as a function of the infrared input power  entering the SFG crystal, while keeping the power of the red light at \SI{770}{\nano \meter} constant at \SI{1}{\watt}. The peak power output is obtained near 3.5\,W.}
	\label{fig:griira}
\end{figure}

The second nonlinear interaction, $\nu + 2\nu = 3\nu$, also employs a MgO:PPLN crystal of 40\,mm length, furnished with a poling period of \SI{6.81}{\micro\meter} and kept at \SI{35}{\degreeCelsius}. With beam foci of around \SI{40}{\micro\meter} for both the 1540-nm and 770-nm lights, up to 600\,mW at 514\,nm are obtained.
At an infrared power of $P_\text{IR}=\SI{10}{W}$, the output power at 514\,nm deviates from the expected $\sim P_\text{IR}^3$ dependence; see Fig.~\ref{fig:power}(b). This divergence is attributed the well-described effect of green-induced infrared absorption (GRIIRA) \cite{GRIIRA}. In the presence of green light, infrared light is increasingly absorbed, which leads to local and inhomogeneous heating of the crystal and thus a decrease of the conversion efficiency. The effect is highlighted in Fig.~\ref{fig:griira}, where the power of generated green light is measured as a function of infrared light entering the SFG crystal, while keeping the power of red light at \SI{1}{W}. Adjusting the power of the infrared light to a level similar to the red light (in our case \SI{3.8}{W}) maximizes the output of frequency converted light. Thus control of the infrared light power before the second MgO:PPLN crystal is required, which is implemented with dichroic mirrors and polarization optics.

For the third conversion stage, $2\nu + 3\nu = 5\nu$, MgO:PPLN is not suitable as it shows high absorption in the UV and would require a very short poling period. Instead, a beta barium borate (BBO) crystal is used in a Type I phase matching configuration with a cutting angle of $\theta = \ang{38}$. The crystal has a length of 10\,mm, the light fields are focused to waists of about \SI{15}{\micro\meter}. A Pellin-Brocca-prism is used to separate the UV light from the other colors, and 0.5\,mW of UV light power are obtained; see Fig.~\ref{fig:power}(c). We measure the power stability at \SI{308}{nm} to be 1.1\% over a time of half an hour.
The frequency-converted light at \SI{308}{nm} is polarised perpendicular to the input beams and thus experiences a walk off inside the birefringent material. This leads to an elliptical output beam, which is beam-shaped using cylindrical lenses.
In the case of BBO, the temperature acceptance bandwidth is about \SI{150}{\degreeCelsius} and thus a temperature stabilisation in the order of a few \si{\degreeCelsius} is sufficient. Here, we measure an optimal temperature of \SI{56}{\degreeCelsius}.\\
We also demonstrated wavelength tunability. To this end, we tuned the output wavelength of the diode laser, and adapt the temperature set-points of the nonlinear crystals accordingly. We also replaced the diode laser by a fiber laser at 1550\,nm, which allowed us to sweep the output wavelength to 310\,nm.

\section{Spectroscopy}
To demonstrate the performance of the laser system for spectroscopy applications, we performe linear Doppler-broadened absorption spectroscopy on the $^1S_0- {}^3P_1$ transition of zinc. This intercombination line has a linewidth of only $2\pi \times 6\,$kHz \cite{Fischer} and is relevant for optical clocks and degenerate quantum gases based on this element. Experiments were performed in a spectroscopy cell, filled with a gas of zinc atoms and heated to \SI{300}{\degreeCelsius}.

\begin{figure}[tb]
	\includegraphics[width=0.95\linewidth]{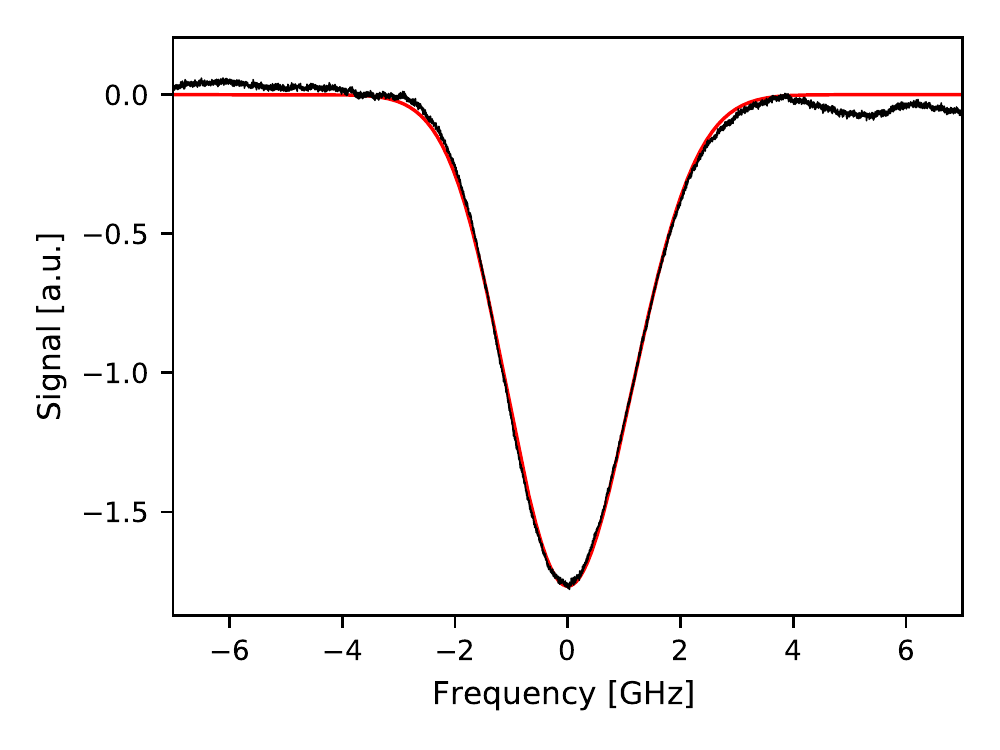}
	\caption{Spectroscopy signal of the $^1S_0- {}^3P_1$ transition in zinc. On the frequency axis the difference to the center frequency of $f = \SI{974 367 479}{MHz}$ is shown. Five overlapping Gaussian functions according to the distribution of natural abundances are fit to the data.}
	\label{fig:spectroscopy}
\end{figure}

The laser light is split into two branches: one of them passes through the spectroscopy cell, the other one serves as a reference to correct for intensity fluctuations via a differencing scheme. We sweep the frequency of the light across the atomic resonance, and record the laser frequency with the help of a wavelength meter (HighFinesse model WS8) with an accuracy of a few MHz. The resulting absorption signal is shown in Fig.~\ref{fig:spectroscopy}. This Doppler-broadened absorption profile is approximated by five overlapping Gaussian functions to account for the distribution of natural abundant isotopes \cite{Gullberg,Campbell}. Both the isotope shifts and the natural abundances are fixed input parameters for the fitting routine; isotope shifts are much smaller than the Doppler-broadened linewidth. The resulting center frequency of the most abundant isotope $^{64}$Zn of

\begin{equation}
f=\SI{974 367 479 (165)}{MHz}
\end{equation}
is consistent, within the uncertainty of the frequency calibration and the fit, with literature values \cite{Gullberg}. Linewidth broadening mechanisms, which are relevant for high-precision spectroscopy, are not studied here. As shown in Ref.~\cite{Herbers}, the relative linewidth $\delta \nu/\nu$ can be maintained in a nonlinear frequency conversion.

\section{Conclusion}
We developed a novel frequency quintupling laser using a three-stage cascaded conversion scheme resulting in UV laser light around \SI{308}{nm}. We achieved a maximum output power of \SI{500}{\micro W}, and demonstrated linear absorption spectroscopy on an intercombination line in zinc. Thus future applications, like performing spectroscopy on the clock transition of zinc or cooling of zinc or beryllium atoms, may be employed using the proposed scheme. To increase the output power, future work will focus on enhancing the conversion efficiency of the last stage.  This could be achieved by using the BBO crystal in a single-resonant \cite{Cheung1994} or double-resonant \cite{Kretschmann1997} bow-tie configuration \cite{Kerdoncuff}. An enhancement of the output power up to the 100-mW level appears feasible and would generate sufficient power for laser cooling applications \cite{Wang_Zn}. With such a resonant enhancement, the alternative approach of frequency quadrupling, $1\nu + 4\nu = 5\nu$, could be revisited.

\section{Backmatter}

\textbf{Acknowledgments} We thank Marcel Hohn for stimulating discussions, and we thank Michael Köhl for generous support and stimulating discussions. We acknowledge funding by Deutsche Forschungsgemeinschaft DFG through grant INST 217/978-1 FUGG and through the Cluster of Excellence ML4Q (EXC 2004/1 – 390534769).\\

\noindent
\textbf{Disclosures} The authors declare no conflicts of interest.\\

\noindent
\textbf{Data availability} Data underlying the results presented in this paper are not publicly available at this time but may be obtained from the authors upon reasonable request.


\bigskip


\printbibliography[]

\end{document}